\input harvmac.tex

\lref\pct{Streater, R.F. and Wightman, A.S.: PCT, spin and
statistics, and all that. Redwood City, USA: Addison-Wesley 1989}

\lref\hawking{Hawking, S.W.: Particle creation by black holes.
Commun. Math. Phys. {\bf 43}, 199-220 (1975) }

\lref\unru{Unruh, W.G.: Notes on black-hole evaporation.
Phys. Rev. {\bf D14}, 870-892 (1976)}

\lref\drinf{Drinfel'd, V.G. and Sokolov, V.V.: Lie algebras and 
equations of Korteweg-de Vries type.
Sov. Math. {\bf 30}, 1975-2036 (1984)}

\lref\reb{Vichirko, V.I. and Reshetikhin, N.Yu.:
Excitation spectrum of the anisotropic
generalization of a $SU_3$ magnet.
Theor. Math. Phys. {\bf 56}, 805-812 (1983)}

\lref\japthree{Jimbo, M., Miki, K., Miwa, T. and  Nakayashiki, A.:
Correlation functions of the XXZ model for $\Delta<-1$.
Phys. Lett. {\bf A168}, 256-263 (1992)}

\lref\daves{Davies, B., Foda. O., Jimbo, M., Miwa, T. and
Nakayashiki, A.: Diagonalization of the $XXZ$ Hamiltonian
by vertex operators. Commun.
Math. Phys. {\bf 151}, 89-153 (1993)}

\lref\jde{Jimbo, J. and Miwa, T.:
Algebraic analysis of Solvable Lattice
Models, Kyoto Univ., RIMS-981 (1994)}

\lref\zamol{Zamolodchikov, A.B.: unpublished}

\lref\lup{Lukyanov, S. and Pugai, Ya.: Multi-point local
height probabilities in the integrable RSOS model.
Nucl. Phys. {\bf B473} [FS], 631-658 (1996)}

\lref\rese{Reshetikhin, N.Yu.:
Diagonalization of $GL(N)$ invariant transfer matrices
and quantum $N$ wave system (Lee model).
Lett. Math. Phys. {\bf 7},
205-213 (1983)\semi
Reshetikhin, N. Yu.: The spectrum of
transfer matrices connected with Kac-Moody
algebras. Lett. Math. Phys. {\bf 14}, 235-246 (1987)}

\lref\kun{Kuniba, A. and Suzuki, J.: 
Analytic Bethe ansatz for fundamental
representations of Yangians. Commun. Math. Phys.
{\bf 173}, 225-264 (1995)\semi
Kuniba, A. and Suzuki, J.: Functional relations and
analytic Bethe ansatz for twisted quantum affine algebras.
J. Phys. {\bf A28}, 711-722 (1995)}

\lref\bazz{Bazhanov, V.V., Nienhuis, B. and
Warnaar, S.O.: Lattice Ising model in a field:
$E_8$ scattering theory, Phys. Lett. {\bf B322},
198-206 (1994)}

\lref\Dod{Dodd, R.K. and Bullough, R.K.: Polynomial
conserved densities for the sine-Gordon equations.
Proc. Roy. Soc. Lond. {\bf A352}, 481-502 (1977)}

\lref\Baxter{Baxter, R.J.: Exactly solved models
in statistical mechanics. London: Academic Press 1982}

\lref\Zhib{Zhiber, A.V. and Shabat, A.B.: Klein-Gordon
equations with a nontrivial group. Sov. Phys. Dokl.
{\bf 24}, 607-609 (1979)}

\lref\Arien{Arinshtein, A.E., Fateev, V.A.
and Zamolodchikov, A.B.:
Quantum S matrix of the\ $(1+1)$-dimensional Todd chain.
Phys. Lett. {\bf B87}, 389-392 (1979)}

\lref\fring{Fring, A., Mussardo, G. and Simonetti, P.:
Form factors of the elementary field in the Bullough-Dodd
model.  Phys. Lett. {\bf B 307}, 83-90 (1993)}

\lref\acerbi{Acerbi, C.: Form factors of 
exponential operators and wave
function renormalization constant in the  Bullough-Dodd
model, Preprint ISAS/EP/2/97, $\#$hep-th/9701062}

\lref\vergeles{ Vergeles, S. and Gryanik., V.: Two-dimensional
Quantum Field Theories having exact solutions.
Yadern. Fiz. {\bf 23}, 1324-1334 (1976)\ (in
Russian) }

\lref\wya{
Awata, H.,
Kubo, H.,
Odake, S. and Shiraishi, J.:
Quantum\ $W_N$\ algebras
and Macdonald Polynomials.
Commun. Math. Phys. {\bf 179}, 401-416 (1996)}

\lref\kar{Karowski, M. and Weisz, P.:
Exact form factors in (1+1)-dimensional
field theoretic models with solution behavior.
Nucl. Phys. {\bf B139}, 455-476 (1978)}

\lref\berg{Berg, B., Karowski, M. and Weisz, P.:
Construction of Green's functions from an
exact S-matrix.
Phys. Rev. {\bf D19}, 2477-2479 (1979)}

\lref\LZ{Lukyanov, S. and Zamolodchikov, A.:
Exact expectation values of local fields
in quantum sine-Gordon model. Nucl. Phys. {\bf B493},
571-587 (1997)}

\lref\ff{
Feigin, B. and Frenkel, E.: Quantum\ $ W$-algebras 
and elliptic algebras.
Commun. Math. Phys. {\bf 178}, 653-678 (1996)}

\lref\Lik{Lukyanov, S.: Form-factors of exponential fields
in the sine-Gordon model. Preprint CLNS 97/1471,
\#hep-th  9703190}

\lref\likk{Lukyanov, S.: Form-factors of exponential fields
in the affine\ $A^{(1)}_{N-1}$\  Toda model. 
Preprint CLNS 97/1478,
\#hep-th  9704213}

\lref\KouM{Koubek, A. and Mussardo, G.:
On the operator content of the sinh-Gordon model.
Phys. Lett. {\bf B311}, 193-201 (1993)}

\lref\ZaZa{Zamolodchikov, A.B. and Zamolodchikov, Al.B.:
Factorized S-matrices in two dimensions as the exact
solutions of certain relativistic 
quantum field theory models.
Ann. Phys. (N.Y.) {\bf 120}, 253-291 (1979) }

\lref\Luk{Lukyanov, S.: Free field representation for massive
integrable models. Commun. Math. Phys. {\bf 167},  183-226 (1995)}

\lref\LUkyan{Lukyanov, S.: A note on the deformed Virasoro
algebra. Phys. Lett. {\bf 367}, 121-125 (1996)}

\lref\yapjj{Jimbo, M., Konno, H. and Miwa T.: Massless
$XXZ$ model and degeneration of the elliptic algebra
$A_{q,p}(\widehat{sl_2})$. Preprint (1996), \#hep-th 9610079 }

\lref\fre{Frenkel, E. and Reshetikhin, N.:
Quantum affine algebras
and deformations of the Virasoro
and\ $W$-algebras. Commun. Math.
Phys. {\bf 178}, 237-266 (1996)}

\lref\yap{
Shiraishi, J., Kubo, H., Awata, H.
and Odake, S.:
A quantum deformation of the
Virasoro algebra and the
Macdonald symmetric functions. Lett. Math. Phys. {\bf 38},
33-51 (1996)}

\lref\lp{Lukyanov, S.  and   Pugai, Ya.: Bosonization
of ZF algebras: Direction toward deformed Virasoro
algebra. JETP {\bf 82}, 1021-1045 (1996)}

\lref\Baxter{Baxter, R.J.: Exactly solved models
in statistical mechanics. London: Academic Press 1982}

\lref\jde{Jimbo, M. and Miwa, T.:
Algebraic analysis of Solvable Lattice
Models. Kyoto Univ., RIMS-981 (1994)}

\lref\daves{Davies, B., Foda. O., Jimbo, M., Miwa, T. and
Nakayashiki, A.: Diagonalization of the $XXZ$ Hamiltonian
by vertex operators. Commun. 
Math. Phys. {\bf 151}, 89-153 (1993)}

\lref\ZamAl{
Zamolodchikov, Al.B.:
Two-point correlation function
in Scaling Lee-Yang model.
Nucl. Phys. {\bf B348}, 619-641 (1991)}

\lref\Lukkk{Lukyanov, S.: Correlators of the Jost
functions  in the Sine-Gordon Model. Phys. Lett.
{\bf B325}, 409-417 (1994)}

\lref\LUkyan{Lukyanov, S.: A note on the deformed Virasoro
algebra. Phys. Lett. {\bf 367}, 121-125 (1996)}

\lref\Fedya{Smirnov, F.A.: Form-factors in completely
integrable models of
quantum field theory. Singapore: World Scientific 1992}

\lref\Mussar{Mussardo, G.:
Off-critical statistical models factorized scattering
theories and bootstrap program.
Phys. Rep. {\bf 218}, 215-379 (1992)}

\Title{\vbox{\baselineskip12pt
\hbox{RU-97-58}
\hbox{CLNS 97/1488}
\hbox{hep-th/9707091}}}
{\vbox{\centerline{
Angular quantization and form-factors }
\vskip6pt
\centerline{in massive integrable models  }}}

\centerline{Vadim Brazhnikov$^{1}$
and Sergei Lukyanov$^{2,3}$}
\centerline{}
\centerline{${}^1$Department of Physics and Astronomy,
Rutgers University}
\centerline{Piscataway, NJ 08855-0849, USA} 
\centerline{${}^2$Newman Laboratory, Cornell University}
\centerline{Ithaca, NY 14853-5001, USA}
\centerline{and}
\centerline{${}^3$L.D. Landau Institute for Theoretical Physics,}
\centerline{Chernogolovka, 142432, RUSSIA}
\centerline{}
\centerline{\bf Abstract}
We discuss an application of the method of
the angular quantization to
reconstruction  of form-factors of local fields
in massive  integrable models. The general formalism
is illustrated  with   examples  of the 
Klein-Gordon, sinh-Gordon  and Bullough-Dodd models.
For the latter two models  the angular quantization approach 
makes it  possible to obtain   free field representations
for form-factors of exponential operators.
We discuss an intriguing relation between the
free field representations and deformations of
the Virasoro algebra. The deformation associated with
the Bullough-Dodd models appears to be different from
the known  deformed Virasoro algebra.

\centerline{}
\centerline{}

\Date{July, 97}
\vfill
\eject

\newsec{Introduction}

The primary goal  of Quantum Field Theory (QFT) is a reconstruction of
a complete set of correlation
functions satisfying a
system of necessary requirements\ \pct.
A way to achieve it is to express  the  correlation functions
in terms of a representation of algebra of local fields
in the  physical Hilbert  space.
The canonical  quantization of Lagrangian
theories provides us with a method to build this representation.
The approach suggests  the following. First of all, we need to choose
some $(D-1)$-dimensional  complete
Cauchy hyper-surface $\Sigma$ in $D$-dimensional space-time. Solution of
underlying classical  system of  relativistically invariant hyperbolic
equations  of  motion  is uniquely defined in the
whole Minkowski space by Cauchy data specified on $\Sigma$. In
the quantum theory  the physical  Hilbert  space appears as a
representation space for canonical  commutation relations defined on
$\Sigma$.
Different choices of the  complete Cauchy hyper-surface lead to unitary
equivalent representations  of the algebra of local fields and do
not affect physically relevant quantities. The
situation changes drastically if we try to quantize the theory using an
incomplete hyper-surface. In this case either there are  no solutions
to the classical system 
of equations of  motion at all or we can reconstruct a
solution only in some region of the
$D$-dimensional  Minkowski space, the so called domain of dependence.
The representation of the algebra of local fields associated with
the incomplete hyper-surface  differs significantly
from the representation in the physical Hilbert
space. It is not clear {\it a priori} what  use  it has
for the purpose of calculation of physically interesting quantities.
The best we might expect is  that QFT can be formulated  in terms of this
``incomplete'' representation with the use of some
density matrix $\hat w$.
It controls  influence  of regions
outside the domain of dependence on the quantum
dynamics. There is no general recipe to
reconstruct $\hat w$.
However in simple cases when the
Cauchy surface is large enough the density matrix can
be recovered from general QFT constraints on
the correlation functions. The most studied
example  corresponds to $D=2$ with a half-infinite 
line $t = 0\, ,\ x>0$ taken
as the
incomplete surface. In this case, solutions of the classical equations
of motion can be defined only in the Rindler wedge,
\eqn\jyrr{x>|t|>0\ .}
The natural parameterization of  
this domain is  provided by the angular
coordinates
\eqn\ksjyrr{x=r\,\cosh(\alpha)\, ,\hskip30pt t=r\,\sinh(\alpha)\ .}
The Hamiltonian picture which appears most natural in this  case is the
one where the polar angle $\alpha$ is treated as time,
and the angular
Hamiltonian ${\bf K}$ generates an infinitesimal shift along the
``time'' direction.
The space of representation of 
the  algebra of local fields associated with
the equal-time slice $\alpha=0$ is usually called by the angular
quantization space. The correlation functions have
to be single valued functions being  continued into the  Euclidean
region. This  
uniquely  
determines the density matrix 
for the angular quantization\ \hawking,
\eqn\densmat{{\hat w }\,=\,e^{2 \pi i {\bf K}}\ .}

The angular quantization turns out to  be a useful
tool to study various aspects of  black
hole evaporation\ \hawking,\ \unru.
There is  strong evidence that it
should be advantageous  for massive integrable models too\ \zamol. 
Typically these models
describe a scaling behavior of exactly solvable 2d statistical
systems. The angular quantization space can be
viewed as the scaling
limit of a space where a  so called corner transfer matrix acts.
Due to R.J. Baxter we know that the latter space possesses
many remarkable  features\ \Baxter. In the  works\ \daves,\ \jde\ 
a nice
algebraic description of them was given. This has  led to  considerable
progress in the calculation of lattice correlation functions\ 
\japthree,\ \jde,\ \lup.
Therefore, it is reasonable to expect that the angular
quantization is  a useful approach to massive integrable QFT.

In this paper we consider an
application of the method of the angular
quantization to reconstruction of form-factors
in integrable QFT containing only one
neutral particle in the spectrum -- the  Klein-Gordon, sinh-Gordon  and
Bullough-Dodd models.

Here is the layout of the paper.
In  Section 2 we briefly review basic features
of  two-body S-matrices and form-factors in integrable QFT. Heuristic
construction\ \Luk\ for the
angular quantization of massive integrable models
is presented in Section 3. In  Section 4
we illustrate the general formalism with
an elementary example of the 
Klein-Gordon model. Section 5 and 6
are devoted to the angular quantization of the sinh-Gordon and
Bullough-Dodd models where we obtain
free field representations for form-factors of
exponential fields. In Section 7 we discuss a relation
between the free field representations and deformations of the Virasoro
algebra. Finally  we conclude with general remarks in Section 8.

\newsec{Preliminaries}

It is a common belief  that the knowledge of 
the two-body  $S$-matrix of integrable models \ZaZa\
makes it  possible, in principle, to
compute  correlation functions of local fields. In order to
elaborate this problem, the so called form-factor 
approach\ \kar, \berg,\ \Fedya\
has been
developed and  successfully employed to study many interesting models.
To simplify our discussion, we will  consider  QFT
with  the  spectrum  
consisting   of a single particle $B$ of mass $m$.
It is convenient to parameterize
the two-body $S$-matrix, describing $B B \to B B$
scattering, in terms of rapidity variables related to
the two dimensional momenta by 
\eqn\lsosiu{p_i^0\,=\,m \cosh(\theta_i)\,,\hskip35pt p_i^1\,=\,m
\sinh(\theta_i)\, ,\hskip35pt i\,=\,1,2\, .} 
The amplitude $S=S(\theta_1-\theta_2)$ satisfies 
general QFT conditions
\eqn\lsskdu{\eqalign{&(i)\,\,\,{\rm unitarity}:\hskip82pt 
S^*(\theta^{*})=S(-\theta)=\big[\, S(\theta)\, \big]^{-1}\cr
&(ii)\,\,\,{\rm crossing\ \  symmetry}:\hskip31pt
S(i\pi-\theta)=S(\theta)\ .\cr}} 
Due to\ \lsskdu\ the amplitude $S(\theta)$ is a $2 \pi i$-periodic
function which is completely determined by the positions of its poles
and zeros in the ``physical strip'' $0<\,\theta<\,\pi$. The simple poles
correspond to ``bound state'' particles either in the direct channel
of $BB$ scattering or in the cross channel, depending on the sign of
the residue.  Since we assume
that there is only one particle in the spectrum the bound state
either does not exist at all 
or must coincide with the  particle $B$ itself
(``$\varphi^3$-property''). In the latter case
the  amplitude $S(\theta)$ develops the
simple pole in the direct channel
at $\theta\,=\,{{2 i \pi} \over {3}}$ 
\eqn\spole{
S(\theta)\,\to\,{ {i\,\Gamma^2}
\over{\theta\,-\,{{2 i \pi} \over {3}}}}   }
and satisfies the bootstrap equation 
\eqn\ficube{S(\theta)= S
\big(\theta-{{i\pi }
\over{3}}\big)\, S\big(\theta\,+\,{ i {\pi }\over{3}}\big)\ .}
We will denote the physical  Hilbert space of the QFT as $\pi_A$.
A linear  basis in $\pi_A$ is  provided by a  set 
of asymptotic states, 
\eqn\kdiduy{|\, B(\theta_n)...B(\theta_1)\, \rangle\ ,}
where rapidities are ordered as $\theta_n>...> \theta_1$.
Any correlation
function can be written as a spectral 
sum over all intermediate $n$-particle
asymptotic states\ \kdiduy.
For example
\foot{Our convention for the normalization of the asymptotic states is
$$ \langle\, vac\, |\, vac\, \rangle=1\,, \ \ \ \ \ \ \
\langle\,   B(\theta)\, |\, B(\theta')
\, \rangle=2\pi\ \delta(\theta-\theta')\ .$$},
\eqn\uyt{\eqalign{&\langle\, {\cal O}(x){\cal O}(y) \, \rangle=
\sum^{\infty}_{n=0}\, \int\limits_{-\infty}^{+\infty}
...\int\limits_{-\infty}^{+\infty}\, {d \theta_1...d \theta_n \over 
n!\  (2 \pi)^n}\
|F_{{\cal O}}(\theta_1,...\theta_n)|^2\ 
e^{-r m \sum^n_{k=1}\cosh\theta_k}\,,\cr
&(x^{\mu}-y^{\mu})^2=-r^2\,<\,0\,.\cr}}
The spectral sum\ \uyt\  involves the on-shell
amplitudes of the local Hermitian field\
${\cal O}$\ (form-factors)
\eqn\hsysr{F_{{\cal O}}(\theta_1,...\theta_n)=\langle\, vac\,|
\, \pi_A ({\cal O})\,  | 
\, B(\theta_n)...B(\theta_1)\, \rangle\ \, , }
where  the  matrix of\ ${\cal O}$\ in the basis of  the asymptotic states
is denoted by $\pi_A ({\cal O})$.
The form-factors satisfy a  set of
requirements\ \kar,\ \Fedya

$(i)$ {\it Watson's theorem,}
\eqn\mars{F_{{\cal O}}
(\theta_1,...\theta_{j+1},\theta_j,...\theta_n)=
S(\theta_j-\theta_{j+1})\ 
F_{{\cal O}}(\theta_1,...\theta_j,\theta_{j+1},...\theta_n)\, .}

$(ii)$ {\it The crossing symmetry condition,}
\eqn\sdv{F_{{\cal O}}(\theta_2,...\theta_{n-1},\theta_n,
\theta_1+2\pi i)=
F_{{\cal O}}(\theta_1,\theta_2,...\theta_{n})\, .}

$(iii)$ {\it The kinematical pole condition.}
$F_{{\cal O}}(\theta_1,...\theta_n)$\
being considered as a function of \ $\theta_n$\
have a simple pole at the point
\ $\theta_n=\theta_{n-1}\,+\,i \pi$\  with the following
residue
\eqn\lsosiu{F_{{\cal O}}(\theta_1,...\theta_{n-1},\theta_n)\,\to\,
 {1-\prod_{j=1}^{n-2} S(\theta_{n-1}-\theta_j)\over
\theta_n-\theta_{n-1}-i \pi}\ \ i\,
F_{{\cal O}}(\theta_1,...\theta_{n-2})\ .}
In presence of the bound state we must supplement the $(i)-(iii)$ with

$(iv)$ {\it The bound state pole condition}. The  
form-factor\ $F_{{\cal
O}}(\theta_1,...\theta_n)$\
being considered as a function of\ $\theta_n$\
have a  simple  pole at 
$\theta_n=\theta_{n-1}\,+\,{{2 i\pi}\over{3}}$ , 
\eqn\boundst{ F_{{\cal O}}(\theta_1,...\theta_{n-1},\theta_n) \,\to\,
{i\, {\Gamma \,F_{{\cal
O}}\big(\theta_1,...\theta_{n-2}, \theta_{n-1}+{i\pi\over 3}
\big)}\over{\theta_n\,-\,\theta_{n-1}\,-\,{{2 i
\pi}\over {3}}}}\,\, ,}
where $\Gamma$ is defined by\ \spole .

\newsec{Heuristic construction}

{}From the mathematical  standpoint,
the system of the form-factor requirements is a complicated
Riemann-Hilbert problem.
The  free field representation approach provides
a technique for its solution.   
It is  based on the following heuristic  construction\ \Luk.

To any factorizable scattering theory there  corresponds
a formal Zamolodchikov-Faddeev (ZF) algebra.
Let us assume that we have some particular  representation $\pi_Z$
of the formal ZF algebra associated with the 
S-matrix $S(\theta)$\ \lsskdu. Denote the ZF operator,  acting in
$\pi_Z$,  by ${\bf B}(\theta)$    
\eqn\uytras{{\bf B}(\theta_1) {\bf B}(\theta_2)=
S(\theta_1-\theta_2)\  {\bf B}(\theta_2) {\bf B}(\theta_1)\
,\hskip30pt \Im m\, (\theta_1-\theta_2)\,=\,0\, .}
We require that the
operator  product ${\bf B}(\theta_2) {\bf B}(\theta_1)$
has a simple pole at the point $\theta_2-\theta_1=i\pi$ with
$c$-number (not operator valued) residue
\eqn\ksisuydy{ {\bf B}(\theta_2) {\bf B}(\theta_1)\,\to\, {i\over 
\theta_2-\theta_1-i\pi}\, . }
In the case of the scattering theory 
exhibiting the  $\varphi^3$-property
it is also required that the operator
product ${\bf B}(\theta_2) {\bf B}(\theta_1)$
develops the simple pole
at  $\theta_2-\theta_1\,=\,{{2 i \pi} \over {3}}$
\eqn\ficubepole{{\bf B}(\theta_2){\bf B}(\theta_1)\,\to\,{i\, {\Gamma
\ {\bf B}(\theta_1 + {{i \pi}\over{3}})}\over{\theta_2\,
-\,\theta_1\,-{{2 i \pi} \over {3}}}}\, .} 
Suppose also there exists an operator \ ${\bf K}$\ acting  in the
space $\pi_Z$ in the following manner
\eqn\bcvcl{{\bf B}(\theta+\alpha)=
e^{-\alpha {\bf K}}\, {\bf B}(\theta)\, e^{\alpha {\bf K}}\ .}
Due to the $S$ matrix properties  \lsskdu, we can impose the
conjugation condition
\eqn\ksisyuty{{\bf B}^+(\theta)=
{\bf B}(\theta+i\pi)\ ,\hskip30pt {\bf K}^+=
-{\bf K} \ .}

Let\ ${\bar \pi_Z}$\ be a dual space to $\pi_Z$.
There exists an embedding of the linear space  of asymptotic
states  $\pi_A$\ in the tensor product
\eqn\hsyt{\pi_A \,\hookrightarrow \, {\bar \pi_Z}\otimes \pi_Z\ .}
In other words, we can identify  an arbitrary  vector
$|\, X \,  \rangle\in \pi_A$\ with
some endomorphism  (linear operator) ${\bf  X}$ of
the space\ $\pi_Z$.
To describe the embedding,  we identify an arbitrary vector $|\,
B(\theta_n)...B(\theta_1)\, \rangle\in\pi_A$ with an element of
${\rm End}\big[\pi_Z\big]$ as 
\eqn\lssoij{|\, B(\theta_n)...B(\theta_1)\, \rangle \equiv
{\bf B}(\theta_n)...{\bf B}(\theta_1)\  e^{i \pi  {\bf K}}\ .}
The  asymptotic states generate the basis
in\ $\pi_A$, therefore\ \lssoij\ unambiguously specifies the
embedding of the linear space. As well as\ $\pi_A$,\ the
space\ ${\bar \pi_Z}\otimes \pi_Z$\ possesses  a  canonical 
Hilbert space structure with the scalar product
given by
$${\rm Tr}_{\pi_Z}\Big[\,
{\bf Y}^+{\bf X}\, \Big]
/{\rm Tr}_{\pi_Z}\Big[ e^{2 \pi i  {\bf K}}\, \Big]\ .$$
We propose the basic conjecture:

{\it 
Eq.\lssoij\ defines
the embedding of the linear space $\pi_A \,\hookrightarrow \, {\bar
\pi_Z}\otimes \pi_Z$, which    preserves  the 
structure of the  Hilbert spaces, }
\eqn\jsusyt{\langle\, Y\, |\, X \,  \rangle={\rm Tr}_{\pi_Z}\Big[\,
{\bf Y}^+{\bf X}\, \Big]
/{\rm Tr}_{\pi_Z}\Big[ e^{2 \pi i  {\bf K}}\, \Big]\, ,\ \ 
\ \ {\rm if}\ \ 
 |\, X \,  \rangle\equiv {\bf X}\, , \ \ 
|\, Y \,  \rangle\equiv {\bf Y}\ .} 
Let us define $\pi_{Z}({\cal O})\in {\rm End}\big[\pi_Z\big]$ 
 associated with the state 
$\pi_{A}({\cal O})|\, vac \, \rangle$ in the following way
\eqn\mvnbb{ \pi_{A}({\cal O})|\, vac \, \rangle
\equiv  \pi_{Z}({\cal O}) 
\ e^{\pi i {\bf K}}\ .}
Notice that for a Hermitian operator ${\cal O}$
\eqn\jdudt{\big[\, \pi_{Z}({\cal O})\, \big]^+=\pi_{Z}({\cal O})\ ,}
as a consequence of our  conjecture.  
Using \ \jsusyt\
the form-factors can be written  as traces
over the space $\pi_Z$,
\eqn\traaaa{F_{{\cal O}}(\theta_1,...\theta_n)=
{\rm Tr}_{\pi_Z}\Big[\, e^{2\pi i {\bf K}}
\ \pi_{Z}({\cal O})\ {\bf B}(\theta_n)...{\bf B}(\theta_1)\, \Big]
/{\rm Tr}_{\pi_Z}\Big[ e^{2 \pi i {\bf  K}}\, \Big]\ .}
The functions\ \traaaa\ satisfy  Watson's theorem\ \mars\  
due to the ZF
commutation relation\ \uytras.    
Using the cyclic property of the trace and the  commutation
relation\ \ksisyuty, we can transform\ \traaaa\ 
$$\eqalign{ \traaaa\,=\,&\,  
{\rm Tr}_{\pi_Z}\Big[\, {\bf B}(\theta_1)\, 
e^{2\pi i {\bf K}}
\ \pi_{Z}({\cal O})\ {\bf B}(\theta_n)...{\bf B}(\theta_2)\, \Big]
/{\rm Tr}_{\pi_Z}\Big[ e^{2 \pi i {\bf  K}}\, \Big]=\cr &\, 
{\rm Tr}_{\pi_Z}\Big[\, 
e^{2\pi i {\bf K}}\  {\bf B}(\theta_1+2\pi i)\ 
\pi_{Z}({\cal O})\ {\bf B}(\theta_n)...{\bf B}(\theta_2)\, \Big]
/{\rm Tr}_{\pi_Z}\Big[ e^{2 \pi i {\bf  K}}\, \Big]\ .}$$
According to the crossing symmetry condition
\sdv, it should coincide with
$$F_{{\cal O}}(\theta_2,...\theta_{n},\theta_1+2\pi i)=
{\rm Tr}_{\pi_Z}\Big[\, 
e^{2\pi i {\bf K}}\ \pi_{Z}({\cal O})\ 
{\bf B}(\theta_1+2\pi i)
{\bf B}(\theta_n)...{\bf B}(\theta_2)\, \Big]
/{\rm Tr}_{\pi_Z}\Big[ e^{2 \pi i {\bf  K}}\, \Big]\ . $$
Hence we conclude  that
$\pi_{Z}({\cal O})\in {\rm End}
\big[\pi_{Z}\big]$ associated  with a local Hermitian field
must commute with ${\bf B}(\theta)$,  
\eqn\lor{\big[\, \pi_{Z}({\cal O})\,  ,\, 
 {\bf B}(\theta)\, \big]=0\ .}
Similarly, basing on the requirements\ \uytras-\bcvcl\ and the  cyclic
property of  a matrix trace, one  can check that the
functions\ \traaaa\ obey the  kinematical pole\ \lsosiu\ and
the bound state pole conditions\ \boundst  . 

Of course our construction
is not a rigorous   method to solve  
the Riemann-Hilbert problem. 
We did not give  a  satisfactory
definition of the representation\ $\pi_Z$\ of the ZF
algebra and a proof of the  basic
conjecture\ \jsusyt. As a result
we can not  prove that the traces\ \traaaa\ are well defined and
satisfy the necessary analytical requirements  imposed  upon 
form-factors. In fact,  a  careful analysis
suggests that  the traces\ \traaaa\ suffer from ultraviolet
divergences and need to be regularized.
Fortunately, these divergences can be  extracted by 
rewriting \traaaa\ in the form 
\eqn\rewwqe{F_{{\cal O}}(\theta_1,...\theta_n)=
\langle \, {\cal O} \, \rangle \
{\rm Tr}_{\pi_Z}\Big[\, e^{2\pi i {\bf K}}
\ \pi_{Z}({\cal O})\ {\bf B}(\theta_n)...{\bf B}(\theta_1)\, \Big]
/{\rm Tr}_{\pi_Z}\Big[ e^{2 \pi i  {\bf  K}}\ 
\pi_{Z}({\cal O}) \, \Big]\ ,}
with
\eqn\hsgst{\langle \, {\cal O} \, \rangle=
{\rm Tr}_{\pi_Z}\Big[ e^{2 \pi i  {\bf  K}}\ 
\pi_{Z}({\cal O}) \, \Big]/{\rm Tr}_{\pi_Z}\Big[ e^{2 \pi i  {\bf  K}}
\, \Big]\ . }
Then  the second factor in\ \rewwqe\ is well defined  and all
divergences are absorbed in the $\theta$-independent
constant $\langle \, {\cal O} \, \rangle$. The latter
is  a  vacuum  expectation value of the local field
${\cal O}$\ \ZamAl. Standard  arguments suggest that
\eqn\ksiyutytr{ \langle \, {\cal O}
 \, \rangle \sim (m \varepsilon )^{d_{{\cal O}}}\ }
up to some finite dimensionless constant.
Here\ $ \varepsilon$\  is  a parameter of the short distance
cut-off (lattice scale) and $d_{{\cal O}}$ is the
ultraviolet scaling  dimension
of the field\ ${\cal O}$. Handling of 
the divergences\ \ksiyutytr\ demands a  proper 
normalization of the local fields\ \ZamAl,\ \LZ. We will not discuss
this problem  here.

\newsec{Klein-Gordon model}

Before  proceeding to analysis of a nontrivial QFT let us demonstrate
how the formal construction from the previous section works for  the
Klein-Gordon model. In this case  the scattering theory is trivial
\eqn\jshdtre{S_{KG}(\theta)=1}
and it is easy to find the proper
representation $\pi_Z$ of the  ZF algebra\ \Lukkk.
Let us introduce a set of  oscillators\ $b_{\nu}$,\ 
satisfying the commutation relations
\eqn\hsfgd{[\, b_{\nu}\, ,\, b_{\nu'} ]
=2 \sinh(\pi\nu)\ \delta(\nu+\nu')\ .}
The algebra admits  a representation  in the Fock space
\eqn\lsjs{{\cal F}:\   \oplus\,  
b_{-\nu_1}...b_{-\nu_n}|\, 0\, \rangle \,  ,\ \  \
\nu_k>0\ .}
The highest vector\ $|\, 0\, \rangle$
(not to be confused with the physical vacuum\ $ |\, vac\, \rangle$)
obeys the equations\
$b_{\nu} |\, 0\, \rangle\,=\,0\, ,\,\,\,  \nu>0.$
The Heisenberg algebra \ \lsjs\
is compatible with  the conjugation
\eqn\ksisy{b_{\nu}^+=b_{-\nu}\ .}
It is easy to check that the  operators
\eqn\jsust{
{\bf B}(\theta)=
\int_{-\infty}^{+\infty} d\nu \ b_{\nu}\  e^{i\nu (\theta-
i{\pi\over 2})}\ ,}
\eqn\ksisudr{{\bf K}=i \int_{0}^{+\infty} {d\nu\over 2}\ {\nu\over
\sinh(\pi \nu)}\ b_{-\nu}b_{\nu} }
satisfy the necessary requirements.
For example,
$$\eqalign{\big[\, {\bf B}(\theta_1)\,  &,
\, {\bf  B}(\theta_2)\, \big]=\int_{-\infty}^{+\infty}\int_{-\infty}
^{+\infty}
d\nu d\nu'\ [\, b_{\nu}\, ,\, 
b_{\nu'}]\  e^{i\nu \theta_1+i\nu' \theta_2}=\cr
&\int^{+\infty}_{-\infty} d\nu\  (e^{\pi\nu }-e^{-\pi\nu }) \
e^{i\nu (\theta_1-\theta_2)}=
2\pi\  \big\{\, \delta(\theta_1-\theta_2-i\pi)-
\delta(\theta_1-\theta_2+i\pi)\, \big\}\ .}$$
Therefore for real $\theta$\ 
\eqn\jssuy{{\bf B}(\theta_1) {\bf B}(\theta_2)=
{\bf B}(\theta_2) {\bf B}(\theta_1)\ .}  
Similar calculations allow one  to check that
$$\eqalign{
& {\bf B}(\theta_2) {\bf B}(\theta_1)\to {i\over
\theta_2-\theta_1-i\pi}\ \ \ \
\ \ {\rm as}\ \ \ \ \theta_2\to\theta_1+i\pi\, 
, \cr 
&{\bf B}(\theta+\alpha)=
e^{-\alpha{\bf  K}}\, {\bf B}(\theta)\, e^{\alpha {\bf K}}  }$$
and the conjugation\ \ksisy\ leads to
$${\bf B}^+(\theta)={\bf B}(\theta+i\pi)\ ,\ \ \ \ \ 
{\bf K}^+=-{\bf K}\ .$$
Thus $\pi_Z$ coincides  with
the Fock space\ ${\cal F}$\ \lsjs.
We expect that
this representation corresponds to  the Klein-Gordon model,
\eqn\sine{(\partial_t^2-\partial_x^2)\, \varphi+m^2\varphi=0\ .}
According to the  discussion from the Section 3,
we should identify  the asymptotic  states
in\ \sine\ with the following endomorphisms of the Fock space,
$$|\, B(\theta_n)...B(\theta_1)\, \rangle \equiv
{\bf B}(\theta_n)...{\bf B}(\theta_1)\  e^{i \pi  {\bf K}}\ .$$
At this point we need to find  the  endomorphism\ $\pi_Z
\big (\varphi(x,t)\big)$ corresponding to the
state
$\pi_{A}\big (\varphi(x,t) \big)|\, vac\, \rangle$.
It can be easily done since
\eqn\hsystr{\pi_{A}\big( \varphi(x,t) \big)|\, vac\, \rangle=
\int _{-\infty}^{+\infty}{{d \theta}\over 
{\sqrt{\pi}}}\  e^{-i m ( x \sinh\theta-t \cosh \theta)}\ 
|\, B(\theta) \, \rangle\  .}
Then  according to our rules\ \lssoij,\ \mvnbb
\eqn\vsfr{\pi_{A}\big( \varphi(x,t) \big)|\, vac\, \rangle\equiv \pi_Z
\big(\varphi(x,t)\big)\ e^{i\pi  {\bf K}}=
\int _{-\infty}^{+\infty}{{d \theta}\over 
\sqrt{\pi}}\  e^{-i mr  \sinh(\theta-\alpha)}\  {\bf B}(\theta)\
e^{i\pi  {\bf K}}\ .}
Here we assume, that\  
$x>|t|>0$\  
and introduce the angular  coordinates\ \ksjyrr\ in the Rindler
wedge. 
Now, let us combine\ \jsust\ with\ \vsfr.  Deforming  the 
integration contour over $ \theta$, we obtain
\eqn\ystre{\pi_Z \big(\varphi(x,t)\big)={2\over \sqrt{\pi}}\ 
\int_{-\infty}^{\infty} d \nu\  b_{\nu}\  e^{i\nu\alpha}\ 
K_{i\nu}(mr)\ ,}
where
$$K_{i\nu}(mr)={1\over 2}\ 
\int _{-\infty}^{+\infty}{d \theta} 
\  e^{- mr  \cosh(\theta)}\ e^{i\nu\theta}\ $$
is  the Macdonald (modified
Bessel) function of the  imaginary order. As  was
expected
\eqn\hsytr{\Big[\pi_Z
\big(\varphi(x,t)\big)\Big]^+=\pi_Z
\big(\varphi(x,t)\big)\, , \ \ \ \
\big[\,  \pi_Z\big(\varphi(x,t)\big)\, ,\,  {\bf B}(\theta)\, \big]=0\ .} 
The formula\ \ystre\ gives a
general solution to the Klein-Gordon equation in 
the Rindler wedge\ \jyrr, satisfying the
canonical commutation relation,
\eqn\hdy{\eqalign{\pi(r)=&
\partial_{\alpha}\varphi(x,t)|_{\alpha=0}\, ,
\ \ \ \varphi(r)=
\varphi(x,t)|_{\alpha=0}\, ,\cr
&[\pi(r),\varphi(r')]=-8\pi i\, r\, \delta(r-r')\  .}}
In this  way we arrive at the 
interpretation of the space\ $\pi_Z$\ as the  angular quantization space
of the model. In this
quantization procedure the angle $\alpha$ is treated
as the ``time'',  
and the  space $\pi_Z$ is associated with
the ``equal time'' slice $\alpha=0$. It is important
that this is a model  independent interpretation of $\pi_Z$.
Moreover the identification of  the physical
vacuum state\ $|\, vac\, \rangle$\  with
$e^{\pi i{\bf K}}\in {\rm End}\big[\pi_Z\big]$
do not suggest
any integrability condition 
and has long been known\ \hawking,\ \unru.

Finishing the elementary example, it is instructive to illustrate
the formula for form-factors\ \traaaa.
According to\ \traaaa,
\eqn\hsyre{\eqalign{
\langle\, vac\, |\, \pi_{A}\big(\varphi(x,t)\big) \, |&
\, B(\theta)\, \rangle
={\rm Tr}_{{\cal F}}\Big[\, e^{2\pi i {\bf K}}\  
\pi_{Z}\big(\varphi(x,t) \big)\ {\bf B(\theta)}\,
\Big]/{\rm Tr}_{{\cal F}}\Big[ e^{2 \pi i  {\bf K}}\, \Big]=\cr
&{2\over \pi}\int_{-\infty}^{+\infty}\int_{-\infty}^{+\infty}
d \nu d\nu'\ 
e^{i\nu\alpha +i\nu'(\theta-{i\pi\over 2})}\
K_{i\nu}(mr)\  
\langle\langle\,  b_{\nu}\, b_{\nu'}\, \rangle\rangle\ ,}}
where
\eqn\jsust{\langle\langle\,  b_{\nu}\, b_{\nu'}\,  \rangle\rangle
={\rm Tr}_{{\cal F}}\Big[\, e^{2\pi i {\bf K}}\
b_{\nu}\, b_{\nu'}\, \Big]
/{\rm Tr}_{{\cal F}}\Big[ e^{2 \pi i  {\bf K}}\, \Big]\ .}
The trace $\langle\langle\,  b_{\nu}\, b_{\nu'}\,  \rangle\rangle$    
can be derived only from  the  cyclic property and
commutation relations for oscillators given in\ \hsfgd
$$\langle\langle\,  b_{\nu}\, b_{\nu'}\,  \rangle\rangle=
e^{2\pi\nu} \langle\langle\, b_{\nu'} b_{\nu}\rangle\rangle=
e^{2\pi\nu} \big\{\, \langle\langle\,  
b_{\nu}\, b_{\nu'}\,  \rangle\rangle-
2\sinh(\pi \nu)\, \big\}\ .$$
Hence
\eqn\hsgre{\langle\langle\,  b_{\nu}\, b_{\nu'}\,  \rangle\rangle=
e^{\pi\nu}\ \delta(\nu+\nu')\ }
and
\eqn\erdtse{
\langle\, vac\, |\, \pi_{A}
\big(\varphi(x,t)\big) \, |\, B(\theta)\, \rangle={2\over \sqrt{\pi}}\, 
\int_{-\infty}^{+\infty}
d \nu\,
e^{-i\nu(\theta-\alpha+i{\pi\over 2})}\
K_{i\nu}(mr)=2\sqrt{\pi}\, 
e^{-i m r\ \sinh(\theta-\alpha)}\ ,}
which is in agreement with\ \hsystr. 
It is important that Wick's theorem can be used
for   calculations of  traces over the free Fock space. The trace
\jsust\ plays the role of Wick's pairing.
By this means the calculation of form-factors of  an arbitrary  local
operator is reduced to a  combinatoric procedure.

\newsec{Sinh-Gordon model}

At the first glance the free field representation for
the angular quantization space appeared
during discussion of the Klein-Gordon model is an artifact
of this particular QFT.
As a matter of fact, the similar free field representations
exist for nontrivial integrable models.
We would like to demonstrate
this important phenomena  on the examples of
the sinh-Gordon and Bullough-Dodd models.

The sinh-Gordon QFT describes dynamics of the  real scalar
field\ $\varphi$\ governed by the Euclidean action,
\eqn\hasy{
{\cal A}_{shG} = \int
d^2 x
\bigg\{\, {1\over {16\pi}}
\big(\partial_{\nu}\varphi\big)^2 +
2\mu\,\cosh\big(b\varphi\big)\, \bigg\}\ . }
There is only one particle in the spectrum which  does not form
any bound states. The corresponding two-body S-matrix does not
have singularities in the physical strip. It was found in the
work\ \vergeles\ that
\eqn\ksustr{S_{shG}(\theta)=
{\tanh({\theta\over 2}- {i\pi b\over 2 Q })\over
\tanh({\theta\over 2}+ {i\pi b\over 2 Q })}\ ,}
where
\eqn\swowjiu{Q=b^{-1}+b\ .}
To describe the free field representation of the angular
quantization space\ $\pi_Z$\ of\ \hasy, let us introduce
the Heisenberg algebra
\eqn\ksiyt{[\, \lambda_{\nu}\, ,\,  \lambda_{\nu'}]=
{2\sinh\big({\pi b\nu\over 2 Q}\big)\,
\sinh\big({\pi \nu\over 2 Q b}\big)\over  \nu\,
\cosh\big({\pi \nu\over 2 }\big)}\ \delta(\nu+\nu')\ }
and the   vertex operator
\eqn\hsydtr{\Lambda(\theta)=
\ :\exp\Big\{\, - i\int_{-\infty}^{+\infty}d\nu\
\lambda_{\nu}\  e^{i\nu ( \theta-i{\pi\over 2})}\, \Big\}:\ .}
Notice that the  set of oscillators, 
which appear during  consideration of
the Klein-Gordon model, is simply related to $\lambda_{\nu}$ in
the limit $b\to 0$,
\eqn\hsgstr{b_{\nu}={2\over \sqrt{\pi}}\
\cosh\big({\pi \nu\over 2 }\big)\
{\rm lim}_{b\to 0} \,
\big(\,  b^{-1}\, \lambda_{\nu}\, \big)\ .}
The algebra\ \ksiyt\ is represented in the Fock space in much the
same way as\ \lsjs\ is.
We extend the construction
and introduce also the
pair of the canonical conjugate operators (``zero modes''), commuting
with the oscillators $\lambda_{\nu}$,
\eqn\jssht{[{\bf P}, {\bf Q}]=-i\ .}
The extended Heisenberg algebra\ \ksiyt,\ \jssht\
admits the representation in the direct sum of the Fock  spaces,
\eqn\gdtsew{\oplus_{p} {\cal F}_{p}\ ,\ \  {\rm where }
\ \ {\bf P} {\cal F}_{p}= p\, {\cal F}_{p}\ .}
In this  space we define the action of operators
${\bf B}(\theta)$  and ${\bf K}$  in the 
following manner\ \LUkyan,\ \yapjj,\ \Lik
\eqn\gste{{\bf B}(\theta)= -i C_{shG}\ \biggl\{\,
e^{i{\pi {\bf P}\over Q}}\, \Lambda\big(\theta+{i\pi\over 2}\big )-
e^{-i{\pi {\bf P}\over Q}}\, \Lambda^{-1}\big(\theta -{i\pi\over 2} \big)
\, \biggr\}\ ,}
\eqn\usytr{{\bf K}=i\ \int_{0}^{+\infty} d\nu\
{2\,  \nu^2\, \cosh\big({\pi \nu\over 2 }\big)
\over  \sinh\big({\pi b\nu\over 2 Q}\big)\,
\sinh\big({\pi \nu\over 2 Q b}\big)} \ \lambda_{-\nu}
\lambda_{\nu}\ .}
One can check that the operators\ \gste\ satisfy
the requirements\ \uytras,\ \ksisuydy,\ \bcvcl.
The real
constant $C_{shG}$ in\ \gste\
should be chosen in order to match
the  normalization of  the
residue in\ \ksisuydy. We will not use its explicit form here.
The conjugation
\eqn\sdew{\lambda_{\nu}^+=\lambda_{-\nu}\ ,\ \ \ \
{\bf P}^+={\bf P}\ }
leads to the  condition\ \ksisyuty .
Hence we  identify the space of the  angular
quantization\ ${\pi_Z}$\  of the sinh-Gordon model with the direct
sum of the  free Fock spaces\ \gdtsew.

Now, we  can take advantage of the proposed free field
representation to  obtain
form-factors of some  local fields in the model
\hasy. In order to do this we need to find  endomorphisms
in ${\pi_Z}=\oplus_{p} {\cal F}_{p}$,  commuting with
the action of the ZF  operator.
Since
$[{\bf P},\Lambda(\theta)]=0$, 
an   obvious  candidate for the proper endomorphism
is a  projector on the  Fock space ${\cal F}_{p_a}$ with a  given
value
$$p_a=a\ .$$
Let us denote the local field
corresponding to  the projector on
${\cal F}_{p_a}$ as ${\cal O}_a$. Then, we have
\eqn\jsusyt{
\langle\, vac\, 
  |\, \pi_A({\cal O}_a) \, |\,
B(\theta_n)...B(\theta_1)\, \rangle
=\langle \, {\cal O}_a \, \rangle\
{\rm Tr}_{{\cal F}_{p_a}}\Big[\, e^{2\pi i {\bf K}}
\  {\bf B}(\theta_n)...{\bf B}(\theta_1)\, \Big]
/{\rm Tr}_{{\cal F}_{p_a}}\Big[ e^{2 \pi i {\bf  K}}\, \Big]\ .}
As was  mentioned already,
Wick's theorem 
can be applied to calculation of  the traces over
the free Fock space.
If we
know the elementary  Wick's pairing, the formulae\ \gste,
\jsusyt\ reduce the calculating of the form-factors to
a combinatoric procedure.
The pairing  can be
evaluated in the same fashion as in the
case of  the Klein-Gordon model.
The rules for 
reconstruction of the form-factors
appear to be as following\ \Lik:
\eqn\sksj{\eqalign{\langle\, vac\, &
  |\, \pi_A({\cal O}_a) \, |\,
B(\theta_n)...B(\theta_1)\, \rangle=\langle\, {\cal O}_{a}\, \rangle
\ \rho^n
\  \times\cr
&\sum_{\{\sigma_j=\pm\}_{j=1}^{n}}\
 e^{i\pi({1\over 2}- { a \over Q})(\sigma_1+...+\sigma_n)}\
\langle \langle\, \Lambda^{\sigma_n}\big(
\theta_n- \sigma_n{i\pi\over 2}\big)\, ...\,
\Lambda^{\sigma_1}\big(\theta_n- \sigma_1{i\pi\over 2}\big)
\, \rangle \rangle\ .}}
The
averaging
$$\langle \langle\, ...\, \rangle \rangle=
{\rm Tr}_{{\cal F}}\Big[ e^{2 \pi i {\bf  K}}\, ...\, \Big]/
{\rm Tr}_{{\cal F}}\Big[ e^{2 \pi i {\bf  K}}\, \Big]$$
is performed with  use of the formula
\eqn\hsydtrre{
\langle \langle\, 
\Lambda^{\sigma_n}(\theta_n- \sigma_n{i\pi\over 2})\, ...\,
\Lambda^{\sigma_1}(\theta_1- \sigma_1{i\pi\over 2})
\, \rangle \rangle=
\prod_{1\leq k<j\leq n} \
R(\theta_{kj})\, \biggl[\, 1+  (\sigma_j-\sigma_k)\,
{i\, \sin\big({\pi b\over Q}\big)
\over 2 \sinh(\theta_{kj})}\, \biggr]\ ,}
where\
$ \theta_{kj}=\theta_k-\theta_j $\  and  the function\ $R(\theta)$\ for
\ $-2\pi<\Im m\theta<0$\ reads
\eqn\slffojks{R(\theta)=\exp\biggl\{- 2\int_{0}^{\infty}
{dt\over t} \ {  \sinh\big({t b\over 2 Q }\big) \,
\sinh\big({t \over 2 Q b} \big)\over \sinh( t)\,
\cosh\big({t\over 2}\big)}\
\cosh\big( t(1-{i\theta\over\pi})\big)\biggr\}\ .}
It is the  so called minimal
form-factor\ \kar,\ \Fedya.
We should choose the constant\ $\rho$\  in\ \sksj\
according to  the
kinematical pole residue normalization\ \lsosiu,
\eqn\hsstre{\rho=
\Big[\, \sin\big({\pi b\over Q}\big)\, \Big]^{-{1\over 2}}\
\exp\biggl\{\int_{0}^{\infty}
{dt\over t} \ { \sinh\big({t b\over 2 Q}\big) \,
\sinh\big({t \over 2 Q b} \big)\over \sinh( t)\,
\cosh\big({t\over 2}\big)}\ \biggr\}\ .}
One can simplify the expressions of the  first
form-factors,
\eqn\sksjsuy{\eqalign{
&\langle\, vac\,|\,  \pi_A({\cal O}_a)
\, |\,  B\,    \rangle
=\langle\, {\cal O}_a\,  \rangle\  h\,   [a]\   ,\cr
&\langle\, vac\,|\,  \pi_A({\cal O}_a)\, |\,
B(\theta_2)B(\theta_1)
\,     \rangle
= \langle\, {\cal O}_a\,  \rangle\ h^2\,    
[a]^2\,  R(\theta_{12})\, \cr
&\langle\, vac\,|\, \pi_A({\cal O}_a)
\, |\, B(\theta_3)B(\theta_2)B(\theta_1)\,
 \rangle=\langle\, {\cal O}_a\,  \rangle\ h^3\,
[a]\,
\prod_{1\leq k<j\leq 3}R(\theta_{kj})
\times\cr
&\ \ \ \ \ \ \ \ \ \ \ \ \ \ \ \ \ \ \ \ \ \ \ \ \ \ \
\biggl\{ [a]^2+
{x_1x_2x_3\over (x_1+x_2)(x_2+x_3)(x_1+x_3)}\biggr\}\ ,}}
here \ $x_k=e^{\theta_k}\, (k=1,2,3)$ and
$$\eqalign{
&[a]={\sin\big({\pi a\over Q}\big)
\over\sin\big({\pi b\over Q}\big)}\ ,\cr
&h=2\rho \sin\big({\pi b\over Q}\big)\ .} $$
The functions\ \sksj,\  \sksjsuy\ coincide 
with the form-factors of exponential
operators in the sinh-Gordon model 
proposed in the work\ \KouM.
Thus we conclude
\eqn\sytaa{{\cal O}_a=e^{a\varphi}\ .}

\newsec{Bullough-Dodd model}

The construction discussed in this section is very similar to
the sinh-Gordon case.
In order to  avoid additional subscripts
we will denote analogous quantities in this section  by the
same symbols as in the previous one.

As an example of QFT possessing the  $\varphi^3$-property we
consider the Bullough-Dodd model\ \Dod,\ \Zhib.
It is  an integrable QFT
defined by the Euclidean action,
\eqn\asytt{
{\cal A}_{BD} = \int
d^2 x
\bigg\{\, {1\over {16\pi}}
\big(\partial_{\nu}\varphi\big)^2 +
\mu\, \big( e^{{\sqrt2}b\varphi}+2 e^{-{b\over{\sqrt2}}
\varphi}\, \big)\, \bigg\}\ . }
The spectrum of the model consists of
one particle that appears to be   the  bound state of itself.
The corresponding two-body S-matrix was proposed in the work\ \Arien
\eqn\ksisuy{S_{BD}(\theta)={{\rm tanh}\big({\theta\over 2}+
{i\pi\over 3}\big)\, {\rm tanh}\big({\theta\over 2}
-{i\pi b\over 3 Q}\big)\, {\rm tanh}\big({\theta\over 2}
-{i\pi \over 3  Q b}\big)\over
{\rm tanh}\big({\theta\over 2}-
{i\pi\over 3}\big)\, {\rm tanh}\big({\theta\over 2}
+{i\pi b\over 3 Q}\big)\, {\rm tanh}\big({\theta\over 2}+
{i\pi \over 3  Q b}\big)}\ ,}
As before we use the notation\ $Q=b^{-1}+b$. Notice  that for $b=1$
the bound state pole disappears and\
\ksisuy\ reduces to the  sinh-Gordon S-matrix with
the coupling constant $b_{shG}={1\over \sqrt2}$.

The space of the angular quantization of the Bullough-Dodd model
is almost identical in construction   to the sinh-Gordon one.
Introduce the set of oscillators obeying the commutation relations
\eqn\ytaaa{[\, \lambda_{\nu}\, ,\,  \lambda_{\nu'}]=
{4\sinh\big({\pi b\nu\over 3 Q}\big)\,
\sinh\big({\pi \nu\over 3 Q b}\big)
\cosh\big({\pi\nu\over 6}\big)\over  \nu\,
\cosh\big({\pi \nu\over 2 }\big)}\ \delta(\nu+\nu')  }
and the  zero modes ${\bf P, Q}$\ \jssht.
As in the previous case
we build   $\pi_Z$ as
a direct sum of the Fock space $\oplus_p {\cal F}_p$.
The ZF  operators read
\eqn\ksoooly{{\bf B}(\theta)= -i C_{BD} \biggl\{\,
e^{i{\pi {\bf P}\over Q}}\, \Lambda
\big(\theta+{i\pi\over 2} \big)-
e^{-i{\pi {\bf P}\over Q}}\,\Lambda^{-1}
\big(\theta -{i\pi\over 2} \big)+
i\,  {1-b^2\over 1+b^2}
\, :\Lambda\big(\theta+{i\pi\over 6} \big)
\Lambda^{-1}\big(\theta -{i\pi\over 6 }\big):
\, \biggr\} ,}
with
$$\Lambda(\theta)=\ :\exp
\Big\{\, -i\int_{-\infty}^{+\infty}d\nu\
\lambda_{\nu}\  e^{i\nu ( \theta-i{\pi\over 2})}\, \Big\}:\ .$$
$C_{BD}$ is a real constant 
providing the normalization\ \ksisuydy.
The  operator ${\bf K}$ is given by the formula which is similar to\ \usytr.

As well as in the sinh-Gordon model  the projectors
on the Fock space with a given eigenvalue of the  zero mode
${\bf P}$  commute with the  ZF operator. Let us denote again
by ${\cal O}_a$ the
local fields which correspond to the projector
on the space\ ${\cal F}_{p_a}$
with
$$p_a={1\over 6}\, \big(4\sqrt2 \, a-b^{-1}+b\big)\ . $$
Thus we have the one parametric family of the
form-factors given by\ \jsusyt.
The trace calculation is similar  to the  one discussed before.
It is convenient to  formulate the result as
the following prescription,
\eqn\issyst{\langle \, vac\, |
\, \pi_A({\cal O}_a)\, |\,
B(\theta_n)...B(\theta_1)\, \rangle=
\langle \, {\cal O}_a\,   \rangle\     \langle\langle\,
{\cal B}(\theta_n)...{\cal B}(\theta_1)\, \rangle\rangle\ ,}
where
\eqn\nsdre{
{\cal B}(\theta)=
\rho \ \biggl\{\, 
\gamma\ \Lambda\big(\theta+{i\pi\over 2}\big)+
\gamma^{-1}\ \Lambda^{-1}\big(\theta-{i\pi\over 2}\big)
+\kappa\ : \Lambda\big(\theta+{i\pi\over 6}\big)
\Lambda^{-1}\big(\theta-{i\pi\over 6}\big):\, \biggr\}\ ,}
and
$$\eqalign{&\gamma=-i \exp\Big({i\pi
\over 6 Q}(4\sqrt2 a-b^{-1}+b)\Big)\ ,\cr
&\kappa=2\, \sin\Big({\pi\over 6 Q} (b^{-1}-b)\Big)\ ,\cr
&\rho=\biggl[\, 
{\sin\big({\pi\over 3}\big)\over
\sin\big({2\pi b\over 3 Q}\big)
\sin\big({2\pi\over 3 Q b}\big)}\, \biggr]^{{1\over 2}}\ 
\exp\biggl\{ 2\int_{0}^{\infty}
{dt\over t} \ 
{\cosh\big({t\over 6}\big)\, \sinh\big({t b\over 3 Q }\big) \,
\sinh\big({t \over  3 Q b} \big)\over \sinh( t)\,
\cosh\big({t\over 2}
\big)}\big)\biggr\}\ .}$$
The  Wick's averaging of the products of the  vertex operators should be
performed
using  the rules
\eqn\sksju{\eqalign{
&\langle\langle\, \Lambda(\theta)\,  \rangle\rangle=1\ ,\cr
&\langle\langle\, \Lambda^{\sigma_2}(\theta_2)\Lambda^{\sigma_1}
(\theta_1) \,  \rangle\rangle=
\Big[R(\theta_1-\theta_2)\Big]^{\sigma_1 \sigma_2}\ , \ \ \
\sigma_i=\pm 1\ ,}}
where
\eqn\ojks{R(\theta)=\exp\biggl\{- 4\int_{0}^{\infty}
{dt\over t} \ {\cosh\big({t\over 6}\big)\, 
\sinh\big({t b\over 3 Q }\big) \,
\sinh\big({t \over  3 Q b} \big)\over \sinh( t)\,
\cosh\big({t\over 2}
\big)}\
\cosh\big( t(1-{i\theta\over\pi})\big)\biggr\}\ }
is the  minimal form-factor for
the Bullough-Dodd model\ \fring.
The  dots\ $:...:$  in\ \nsdre\ mean
that we do not
need to pair the  vertex operators within
the normally ordered group under
Wick's averaging.
{}From\ \issyst-\ojks, one can easily derive the first two form-factors
\eqn\jsbcgfdr{\eqalign{&\langle\, vac \, |\,
\pi_A( {\cal O}_a)\, | \, B\,  \rangle=
\langle \, {\cal O}_a\,
\rangle \ \rho\, (\gamma+\gamma^{-1}+\kappa)\ ,\cr
&\langle \, vac\, |\, \pi_A( {\cal O}_a)
\, | \, B(\theta_2)B(\theta_1)\,  \rangle=
\langle \, {\cal O}_a\,\rangle \ \rho^2 \ R(\theta_{12})
\Big(\,   \gamma^2+\gamma^{-2}
+G\big(\theta_{12}-{i\pi\over 2}\big)+\cr
&\ \ \ \ \ \ \ \ \ G\big(\theta_{12}+
{i\pi\over 2}\big)+
\kappa (\gamma+\gamma^{-1})
\big(\, F(\theta_{12}-{i\pi\over 3})+F(\theta_{12}+
{i\pi\over 3})\, \big)+
\kappa^2  F(\theta_{12})\,      \Big)\ .}}
Here we use the notations,
\eqn\jsudt{\eqalign{&F(\theta)={R(\theta)\over R(\theta-{i\pi\over 3})
R(\theta+{i\pi\over 3})}\ ,\cr
&G(\theta)={1\over R(\theta-{i\pi\over 2}) R(\theta+{i\pi\over 2})}\ .}}
The functions $F$ and $G$ read explicitly,
\eqn\jsudyt{\eqalign{&F(\theta)={\sinh\big({\theta\over 2}+
{i\pi\over 6 Q} (b^{-1}-b)\big)\,
\sinh\big({\theta\over 2}-
{i\pi\over 6 Q} (b^{-1}-b)\big)\over
\sinh\big({\theta\over 2}+{i\pi\over 6}\big)\,
\sinh\big({\theta\over 2}-{i\pi\over 6}\big)} ,\cr
&G(\theta)={\sinh\big({\theta\over 2}-{i\pi\over 4}
+{i\pi b\over 3 Q}\big)
\sinh\big({\theta\over 2}-{i\pi\over 4}
+{i\pi \over 3 Q b}\big)
\sinh\big({\theta\over 2}+{i\pi\over 4}
-{i\pi b\over 3 Q}\big)
\sinh\big({\theta\over 2}+{i\pi\over 4}
-{i\pi \over 3 Q b}\big)\over
\sinh\big({\theta\over 2}+{i\pi\over 4}\big)\,
\sinh\big({\theta\over 2}-{i\pi\over 4}\big)\,
\sinh\big({\theta\over 2}+{i\pi\over 6}\big)\,
\sinh\big({\theta\over 2}-{i\pi\over 6}\big)}
\ .}}
We can  simplify\ \jsbcgfdr\ and present in the form
\eqn\ksisy{\eqalign{&\langle \, vac\, |\,
\pi_A( {\cal O}_a)\, | \, B\,  \rangle=
\langle \, {\cal O}_a\,\rangle\   h\,
\{a\}\ ,\cr
&\langle \, vac\,|\,
\pi_A(  {\cal O}_a)\, | \, B(\theta_2)B(\theta_1)\,  \rangle=
\langle \, {\cal O}_a\,\rangle \  h^2\, \{a\}\,
R(\theta_{12})\
\Big(\,  \{a\}-{ x_1x_2\over
x_1^2+x_2^2+x_1x_2}\, \Big)\ ,}}
where $x_k=e^{\theta_k}\ (k=1,2)$ and
$$\eqalign{&\{a\}={\sin\big({\pi \sqrt2 a\over 3 Q}\big)\
\cos\big({\pi \over 6 Q}(2\sqrt2 a-b^{-1}+b)\big)\over
2 \sin\big({\pi\over 6 Q} (b^{-1}-b)\big)
\sin\big({\pi b\over 3 Q}\big) \sin\big({\pi \over 3 Q b}\big)}\   ,\cr
&h=8 \rho
\, \sin\big({\pi\over 6 Q} (b^{-1}-b)\big)
\sin\big({\pi b\over 3 Q}\big) \sin\big({\pi \over 3 Q b}\big)\ .}$$
The functions\ \ksisy\ coincide with the form-factors
of the  exponential
fields  $e^{a\varphi}$  found in the work\ \acerbi. We have also
checked that\ \issyst\ reproduces
the  three particle form-factors
presented in\ \acerbi. 
In all likelihood\ ${\cal O}_a=e^{a\varphi}$\ again.

\newsec{Deformations of  Virasoro algebra}

In this section we would like  
to discuss an intriguing relation between the 
ZF operators for the sinh-Gordon and Bullough-Dodd models and
deformations of the 
Virasoro algebra.
Let  us introduce the real parameter 
$$0<x<1\ .$$
The  ZF
operator ${\bf B}_{shG}(\theta)$\ \gste\ can be written
as\ $x\to1$\  (``scaling'')  limit,
\eqn\shsytr{{\bf B}_{shG}(\theta)=C_{shG}\ {\rm lim}_{x\to 1} \
{\bf U}\big(x^{{2i\theta\over \pi}}\big)\ ,}
where
\eqn\hsyst{\eqalign{&
{\bf U}(z)= -i
\  \biggl\{\,
e^{i{\pi {\bf P}\over Q}}\, \Lambda_x(z x^{-1}) -
e^{-i{\pi {\bf P}\over Q}}\, \Lambda^{-1}_x(z x) 
\, \biggr\}\ , 
\cr &
\Lambda_x(z)=\ 
:\exp\biggl\{\, -i
\sum_{n\not= 0}\ \lambda_{n}\ \big(\, z x\, \big)^{-n}
\, \biggr\}:\ .} } 
The oscillators $\lambda_n$,
\eqn\hsystr{[\, \lambda_{n}\, ,\,  \lambda_{m}]={
\big(x^{ {n b\over Q}}- x^{- {n b\over Q}}\big)
\big(x^{ {n \over Q b}}- x^{- {n \over Q b}}\big)\over
n\, (x^n+x^{-n})}\ \delta_{n+m,0}\ , }
are  specified by   an integer $n$
rather then   continuous parameter $\nu\to - {2n\over \pi}\log( x)\ 
( x\to 1) $ as it was in \hsydtr.
In other words,  we treat the 
Fourier integral in\ \hsydtr\ as a limiting value of
the proper  Fourier series.
The ``deformed'' ZF operator generates a  quadratic algebra with
the following commutation relation\ \fre,\ \yap,\ \ff
\eqn\ustasa{\eqalign{
f\big(\zeta z^{-1}\big)\,  {\bf U}(z){\bf U}(\zeta)-&
f\big(z \zeta^{-1}\big) \, {\bf U}(\zeta) {\bf U}(z)=\cr
&
\big(x-x^{-1}\big)\, \Big\{\, \delta\big(z \zeta^{-1} x^{-2}\big)-
\delta\big(z \zeta^{-1} x^{2}\big)\, \Big\}\
\Big[{ b\over  Q}\Big]_x\Big[
{1 \over  Q b}\Big]_x\ ,}}
with
\eqn\jsusyt{f(z)={\big(z x^{{2 b\over Q}};x^4\big)_{\infty}
\big(z x^{{2 \over Q b}};x^4\big)_{\infty}\over
(1-z) \big(z x^{2+{2 b\over Q}};x^4\big)_{\infty}
\big(z x^{2+{2 \over Q b}};x^4\big)_{\infty}}\ .}
Here we use the conventional  notations,
$$\eqalign{&
\big(z;q\big)_{\infty}=\prod_{n=0}^{+\infty} (1-z q^n)\ ,\cr
&[a]_x={x^a-x^{-a}\over x-x^{-1}}\ ,\cr
&\delta(z)=\sum_{n=-\infty}^{+\infty} z^n\ .}$$
The operator ${\bf U}(z)$ admits  the
power series expansion,
\eqn\hsystr{{\bf U}(z)=\sum_{n=-\infty}^{+\infty}\ 
{\bf U}_n\ z^{-n}\ ,}
and Eq.\ustasa\ can be equivalently rewritten in terms of 
the modes ${\bf U}_n $
\eqn\hsystr{\sum_{k=0}^{+\infty}\ f_k \, 
\big(\, 
{\bf U}_{n-k} {\bf U}_{m+k}-{\bf U}_{m-k} {\bf U}_{n+k}\, \big)=
\big(x-x^{-1}\big)^2 \, 
\Big[{ b\over  Q}\Big]_x\Big[
{1 \over  Q b}\Big]_x\, 
[2n]_x\ \delta_{n+m,0}\ ,}
where the  coefficients\ $f_k$\ are defined  by the formula
$$f(z)= 
\sum_{k=0}^{+\infty}\ f_k\  z^k\ .$$
The algebra\ \hsystr\
is a   deformation of the  famous Virasoro algebra. Indeed,
assuming
\eqn\lssisy{{\bf U}_{n}= 2 \delta_{n,0} - \big(x-x^{-1}\big)^2 
\,   \Big(\, Q^{-2}\,   {\bf L}_n-
{\delta_{n,0}\over 4}\, \Big)+O\Big((x-x^{-1}\big)^4\Big)\ , }
\hsystr\ leads to  the 
well known commutation relations for  
the generators\ $ {\bf L}_n$,
\eqn\nsgdr{[\, {\bf L}_n\, ,\,  {\bf L}_m]=(n-m)\, {\bf L}_{n+m}+
{c\over 12}\ (n^3-n)\ , }
where $c=c(b)$,
\eqn\lsskju{c(b)=1+6\,  \big(\, b+b^{-1}\, \big)^2\  .}

The ZF operator
corresponding to
the Bullough-Dodd model\ \ksoooly\ can be deformed
in  analogous way,
\eqn\shsytr{{\bf B}_{BD}(\theta)=C_{BD}\ {\rm lim}_{x\to 1} \
{\bf V}\big(x^{{2i\theta\over \pi}}\big)\ .}
The ``current'' ${\bf V}(z)$ also  generate the quadratic algebra,
\eqn\jsust{\eqalign{&
g\big(\zeta z^{-1}\big)\,  {\bf V}(z){\bf V}(\zeta)-
g\big(z \zeta^{-1}\big) \, {\bf V}(\zeta) {\bf V}(z)=\cr
&
\big(x-x^{-1}\big)\, \Big\{\, \delta\big(z \zeta^{-1} x^{-2}\big)- 
\delta\big(z \zeta^{-1} x^{2}\big)\, \Big\}\ 
{\big[{2 b\over 3 Q}\big]_x\big[{2 \over 3 Q b}\big]_x
\big[{1\over 3}+{2 b\over 3 Q}\big]_x\big[{1\over 3}+
{2 \over 3 Q b}\big]_x\over \big[{1\over 3}\big]_x \big[{2\over 3}\big]_x}+
\cr
&\big(x-x^{-1}\big)\, \Big\{\, \delta\big(z \zeta^{-1}
x^{-{4\over 3}}\big)\, \big)
{\bf V}\big(zx^{-{2\over 3}}\big)-
\delta\big(z \zeta^{-1} x^{{4\over 3}}\big)
{\bf V}\big(zx^{{2\over 3}}\big)\Big\}\
{\big[{2 b\over 3 Q}\big]_x\big[{2 \over 3 Q b}\big]_x
\big[{b^{-1}-b\over 3 Q}\big]_x
\over \big[{1\over 3}\big]_x \big[{2\over 3}\big]_x}\ ,}}
where
\eqn\trseqw{\eqalign{g(z)=&
{\big(x^{{10\over 3}} z;x^{4}\big)_{\infty}
\big(x^{{8\over 3}} z;x^{4}\big)_{\infty}
\over
(1-z)
\big(x^{{4\over 3}} z;x^{4}\big)_{\infty}
\big(x^{{2\over 3}} z;x^{4}\big)_{\infty}}\times\cr
&{\big(x^{{2\over 3}+{4 b\over 3Q}} z;x^{4}\big)_{\infty}
\big(x^{{2\over 3}+{4\over 3Qb}} z;x^{4}\big)_{\infty}
\big(x^{{4 b\over 3Q}} z;x^{4}\big)_{\infty}
\big(x^{{4\over 3Qb}} z;x^{4}\big)_{\infty}\over
\big(x^{{8\over 3}+{4 b\over 3Q}} z;x^{4}\big)_{\infty}
\big(x^{{8\over 3}+{4 \over 3Qb}} z;x^{4}\big)_{\infty}
\big(x^{2+{4b\over 3Q}} z;x^{4}\big)_{\infty}
\big(x^{2+{4\over 3Qb}} z;x^{4}\big)_{\infty}}\ .}}
In terms of the  Laurent modes,\ 
${\bf V}(z)=\sum_{n=-\infty}^{+\infty}\
{\bf V}_n\ z^{-n},$\ the
commutation relation\ \jsust\ reads as
\eqn\strewqe{\eqalign{\sum_{k=0}^{+\infty}&\ g_k \,
\big(\,
{\bf V}_{n-k} {\bf V}_{m+k}-{\bf V}_{m-k} {\bf V}_{n+k}\, \big)=\cr
&\big(x-x^{-1}\big)^2 \,
{\big[{2 b\over 3 Q}\big]_x\big[{2 \over 3 Q b}\big]_x
\big[{1\over 3}+{2 b\over 3 Q}\big]_x\big[{1\over 3}+
{2 \over 3 Q b}\big]_x\over \big[{1\over 3}\big]_x \big[{2\over 3}\big]_x}\
[2n]_x\ \delta_{n+m,0}+\cr
&\ \ \ \ \ \ \ \ \big(x-x^{-1}\big)^2 \,
\ {\big[{2 b\over 3 Q}\big]_x\big[{2 \over 3 Q b}\big]_x
\big[{b^{-1}-b\over 3 Q}\big]_x
\over \big[{1\over 3}\big]_x \big[{2\over 3}\big]_x}\ 
\biggl[{2(n-m)\over 3}\biggr]_x
\ {\bf V}_{n+m}\  .}}
Here\ $g_k$\ are generated by the function $g(z)$
$$g(z)=
\sum_{k=0}^{+\infty}\ g_k\  z^k\ .$$
Again, assuming that
\eqn\isy{{\bf V}_{n}=\pm v\big(b^{\pm1}\big)\, \delta_{n,0}\mp
\big(x-x^{-1}\big)^2\ \Big(\,   {8\over 9 Q^2}\
{\bf L}^{\pm}_n\, - s\big(b^{\pm 1}\big)\,  \delta_{n,0}\, \Big)
+O\Big((x-x^{-1}\big)^4\Big)\ , }
where
$$\eqalign{&v(b)={3+b^2\over Q b}\ ,\cr
&s(b)={(3+b^2)(8+13 b^2+6 b^4)\over
54\,    Q^3\,  b^3}\ ,}$$
one can check that
\ $ {\bf L}^{\pm}_n$\ obey\ \nsgdr\ with the central charges
$$c^{\pm}=c\big(\sqrt2b^{\mp 1}\big)\ .$$
The function\ $c(b)$\ is given by\ \lsskju. 
We conclude that both\ \hsystr\ and\ \strewqe\
provide  associative
deformations of the Virasoro algebra. They look very different.
It would be interesting to analyze an   equivalence
(or nonequivalence) of these deformations.

\newsec{Conclusion}

Our study of the angular quantization spaces
suggests
that the following  observations are quite common features
of integrable QFT models:

({\it i}) Form-factors of local operators can be represented
as traces over the space of angular quantization.

({\it ii}) The space of the  angular quantization 
for a  massive integrable
model admits description in terms of
free Fock spaces. Therefore
the  traces can be easily evaluated.
 
({\it iii}) The space of the  angular 
quantization for a  massive integrable model
can be treated as a  ``scaling'' limit of a
representation of  some ``deformed'' algebra.
The ZF operators corresponding
to a  diagonal scattering theory  are the  scaling limits
of  currents of the deformed algebra.

The construction,  similar to the one  discussed in the
body of the paper,  has  been 
recently developed for 
the affine $A^{(1)}_{N-1}$ Toda QFT, which 
contains $N-1$ particle in the spectrum\ \likk.
In this case the ZF operators are
the  ``scaling'' limits of  
currents of the deformed $WA_{N-1}$ algebra\ \ff,\ \wya.

We can not resist the temptation to mention here a remarkable
similarity of
the free field representations of  ZF operators and
famous Baxter $T-Q$  equations\ \Baxter
\foot{The resemblance  of the deformed W-algebras and
Baxter equations was already noted by
E. Frenkel and N. Reshetikhin\ \fre.}.
It can be illustrated 
with the example of the  Bullough-Dodd model.
The model belongs to the class of the affine Toda QFT corresponding
to the\ $A^{(2)}_2$\ root system\ \drinf\
while
the Baxter equation
associated with this 
root system reads\ \reb,\ \bazz
\eqn\hsgdra{\eqalign{{\cal T } 
(\theta)\  {\cal Q}\big(\theta+{ i\pi\over 6}\big)\, 
&{\cal Q}\big(\theta-{ i\pi\over 6}\big)=
\gamma(\theta)\
{\cal Q}\big(\theta+{ 5i\pi\over 6}\big)\, 
{\cal Q}\big(\theta-{ i\pi\over 6}\big)+\cr &
\gamma^{-1}(\theta)\  {\cal Q}\big(\theta-{ 5i\pi\over 6}\big)\, 
{\cal Q}\big(\theta+{ i\pi\over 6}\big)+
\kappa(\theta)\  {\cal Q}(\theta+{i\pi\over 2})\, 
{\cal Q}(\theta-{ i\pi\over 2})\ ,}}
where ${\cal T}(\theta)$ and ${\cal Q}(\theta)$ are 
the fundamental transfer matrix and 
Baxter $ Q$-operator,  respectively.
The numerical functions 
$\gamma(\theta)$\ and $\kappa(\theta)$ are non universal 
and depend
on  a specific model from  the $A^{(2)}_2$ class.
If we introduce
$$\Lambda(\theta)={{\cal
Q}\big(\theta+{ i\pi\over 3 }\big)\over
{\cal Q}\big(\theta-{ i\pi\over 3 }\big)}\ ,$$
the form
of the  $T-Q$ 
equation
will  essentially coincide with  the
formula\ \nsdre.
We expect that  the numerous  results on the Baxter
equations
associated with the different root
systems\ \rese,\ \kun\ will be 
extremely useful to develop  the
angular quantization of general  affine Toda QFT.

\hskip1.0cm

\centerline{\bf Acknowledgments}

We are grateful to A.B. Zamolodchikov
for sharing his insights and interesting discussions.
S.L.  acknowledges  helpful discussions with 
V. Korepin and  G. Mussardo
and thanks the Department of Physics and Astronomy, Rutgers University
for the hospitality.
This work 
is supported in part by NSF grant (SL).

\listrefs

\end